\documentclass[pra,preprint,fleqn]{revtex4-1}
\usepackage{graphicx}
\usepackage{amsmath}
\usepackage{amssymb}
\usepackage{amsmath}
\usepackage{color}
\usepackage{bm}
\usepackage{latexsym}
\usepackage{hyperref}
\usepackage{float}
\usepackage{graphicx}
\usepackage{epstopdf}
\usepackage{subfigure}
\newcommand{\void}[1]{}

\newcommand{\be}{\begin{equation}}
\newcommand{\ee}{\end{equation}}
\newcommand{\ket}[1]{\ensuremath{\left|{#1}\right\rangle}}
\newcommand{\bra}[1]{\ensuremath{\left\langle{#1}\right|}}

\newcommand{\R}{\mathbb{R}}
\newcommand{\subps}[3]{\ensuremath{{#1}_{\mbox{\tiny{#2}}}^{\mbox{\tiny{#3}}}}}

\DeclareUnicodeCharacter{2212}{-}

\begin{document}
\title{Quantum approach to the thermalization of the toppling pencil interacting with a finite bath}
\author{Sreeja Loho Choudhury}
\affiliation{Max-Planck-Institut f\"ur Physik Komplexer
Systeme, N\"othnitzer Str. 38, D-01187 Dresden, Germany}
\affiliation{Institut f\"ur Theoretische Physik, Technische Universit\"at
Dresden, D-01062 Dresden, Germany}
\author{Frank Grossmann}
\affiliation{Institut f\"ur Theoretische Physik, Technische Universit\"at
Dresden, D-01062 Dresden, Germany}
\date{\today}

\begin{abstract}
We investigate the longstanding problem of thermalization of quantum systems coupled to an environment
by focusing on a bistable quartic oscillator interacting with a finite number of harmonic oscillators.
In order to overcome the exponential wall that one usually encounters in grid based approaches to solve
the time-dependent Schr\"odinger equation of the extended system, methods based on the
time-dependent variational principle are best suited. Here we will apply the method of coupled
coherent states [D. V. Shalashilin and M. S. Child, J. Chem. Phys. {\bf 113}, 10028 (2000)].
By investigating the dynamics of an initial wavefunction on top of the barrier of the double well,
it will be shown that only a handful of oscillators with suitably chosen frequencies, starting in their ground states,
is enough to drive the bistable system close to its uncoupled ground state. The long-time average of
the double-well energy is found to be a monotonously decaying function of the number of environmental oscillators
in the parameter range that was numerically accessible.

\end{abstract}

\maketitle

\section{Introduction}

Ever since the pioneering works of Fermi, Pasta, Ulam and Tsingou (FPUT) \cite{FPUT65,Ford92}, the puzzle of energy exchange between subsystems, eventually 
leading to equilibration and thermalization in closed systems with a finite number of degrees of freedom has intrigued researchers in the fields of 
classical \cite{BeRo93,Jarz95,BoAg06,RoBe08,SO08,MaAg11,MFA12,Jinetal13,DPM20}, as well as semiclassical \cite{jpca12,CoFi14} and quantum mechanics 
\cite{JeSh85,BeRo93,GMM09,KN09,GBS14,KCV16,Rei16,BoGe20}. In the quantum context, an important concept is the eigenstate thermalization
hypothesis (ETH) \cite{AKPR16,De18}. Whereas thermalization in classical systems is closely related to the presence of chaos
and ergodicity, the ETH can be regarded, very broadly, as the quantum manifestation of such ergodic
behavior. The principal philosophy underlying the ETH, however, is that instead of explaining the ergodicity of
a thermodynamic system through the mechanism of dynamical chaos, as done in classical mechanics,
one should instead examine the properties of matrix elements of observable quantities in individual
energy eigenstates of the system. The eigenstate thermalization hypothesis states that for an arbitrary
initial state of the system, the expectation value of any observable (of a subsystem) will ultimately evolve in time to
its value predicted by a micro-canonical ensemble, and thereafter will exhibit only small fluctuations
around that value.

In classical systems the prerequisite for thermalization is believed to be the (hard) chaoticity of the underlying dynamics \cite{Jarz95,BoAg06};
this fact being one possible reason that by investigating a weakly anharmonic system, FPUT did not succeed in finding thermalization but were surprised by a dynamics
that showed pronounced revivals \cite{Ford92}. For dynamical chaos to appear, the phase space has to be at least three dimensional.
The question therefore arises if it is enough to increase the interaction strength between the different degrees of freedom in order to fully develop chaos,
or if, in addition, the number of those degrees of freedom has to be increased (possibly all the way to the thermodynamic limit) in order to
observe thermalization. In two more recent contributions from the realm of classical mechanics it has been shown by solving Newton's equations
for coupled harmonic oscillator systems, comprising a few hundreds \cite{SO08} up to a few thousand degrees of freedom \cite{Jinetal13} that, 
for a suitably chosen initial configuration, a (large enough) subsystem may indeed reach thermal equilibrium, without coupling to an (external) thermostat
and even without nonlinear interactions (i.e.\ without chaos). In the light of these results, another possible reason that FPUT saw no signs of thermalization
is the closeness of their model to the noninteracting case \cite{Jinetal13}.

Although semiclassical approaches may be helpful to tackle such problems \cite{jcp18}, this large number of degrees of freedom seems elusive
if one is interested in a full quantum description of the process. From this perspective it comes as a relief that also small numbers
of degrees of freedom (below ten) can lead to thermalization in long-time dynamics of quantum systems, as shown long ago for spin systems \cite{JeSh85}
and discussed more recently in the seminal book by Gemmer, Michel and Mahler \cite{GMM09} and the article by 
Reimann \cite{Rei16}. For a more recent reference on spin systems realized in graphene quantum dots, see \cite{HFT15} and for energy exchange in quantum systems 
with continuous degrees of freedom, see \cite{GBS14}. There it is claimed that as little as ten to twenty harmonic degrees of freedom are necessary to observe energy loss to the bath
without backflow on the observed time scales. In addition, in \cite{KN09}, the thermalization of eight valence electrons inside a small sodium cluster has been investigated
quantum mechanically. Furthermore, 
whereas in \cite{JeSh85} it was argued that both integrable as well non-integrable systems exhibit statistical behavior for long times, 
in \cite{KCV16} it was pointed out that, in a Bose-Hubbard dynamics, thermalization is only observed if the system starts from a chaotic region of 
phase space but not if the system is launched from a quasi-integrable region.

In the present contribution, we will investigate the toppling pencil model, studied recently by Dittrich and Pena Mart{\'i}nez (DPM) \cite{DPM20}.
In their classical mechanics study, they focused on a particle on top of the barrier in a quartic double well that is coupled to a small 
(from just one single up to the order of ten and higher)
number of harmonic oscillators. It is well known that the transition from integrable to chaotic motion sets in first around the
separatrix of the double well's  phase space, when the interaction strength with an external sinusoidal field is tuned higher \cite{klu95}. 
In the light of the findings in \cite{KCV16},
this makes an initial condition starting on top of the double well's barrier an ideal candidate to search for (quick \cite{Rei16}) thermalization under 
the interaction with a relatively small number of harmonic degrees of freedom, although the interaction with a sinusoidal field is a crude approximation 
to the interaction with harmonic degrees of freedom, whose dynamics is influenced by ``back action'' of the system.

In contrast to  DPM, in the following, we will investigate the system dynamics fully
quantum mechanically, making use of the method of coupled coherent states, introduced by Shalashilin and Child \cite{SC00}. By being based on an
expansion of the total wavefunction in terms of coherent states (Gaussians), whose initial positions in phase space are chosen randomly, the 
exponential wall that one usually experiences in grid based approaches to the quantum dynamics can be overcome or at least be pushed to rather
larger numbers of degrees of freedom. For a recent review of the CCS and related methods, we refer to \cite{irpc21}. The main focus of the present work 
is on the question of how many bath degrees will be needed to ensure thermalization and
how strong the interaction has to be and how the speed of thermalization depends on the coupling and/or the number of degrees of freedom.
The toppling pencil model setup seems to be ideally suited to answer all those questions.

The manuscript is structured as follows: In Sec.\ II, we briefly introduce the bistable quartic oscillator toppling pencil model coupled to a harmonic heat bath. 
In Sec. III, the CCS method that will be employed to study the quantum dynamics of the many particle system will be reviewed. In Sec.\ IV, numerical results for 
several quantities of interest are presented. These are energy expectation values, auto-correlation
functions, as well as reduced densities that allow us to observe the transition of the quartic degree of freedom into some almost stationary state
resembling the ground state of the unperturbed problem to a large degree. Some conclusions as well as an outlook are given in the final section.
The appendix contains details for the implementation of the CCS method.

\section{Quartic Double Well coupled to a Finite Heat Bath}

Our model system is a double well, which is bilinearly coupled to a finite number $N$ of harmonic oscillators \cite{DPM20}. A quartic double-well
is a bistable system with two symmetry related minima. It has many physical realizations, one of the most prominent of which is the ammonia molecule, first 
discussed in the quantum (tunneling) context by Hund as early as 1927 \cite{Hu27}. A solid state realization of a non quartic but symmetric double-well 
potential is given by a suitably parametrized rf-SQUID, where the role of the coordinate is played by the flux through the ring \cite{Ku72}.
More recently bistable potentials have been discussed in cold atom physics in connection with Bose-Einstein condensation \cite{Ketal08}.
In the following, we first discuss the bare quartic bistable system before coupling it to a finite heat bath.

\subsection{Quartic Double Well}
\label{ssec:DW}

The potential of a symmetric quartic oscillator double well with a parabolic barrier around its relative maximum can be written as
\be
V_{S}(x)=-\frac{a}{2}x^{2}+\frac{b}{4}x^{4},\qquad
a, b\in \R^{+} .
\ee
It has quadratic minima at $x_{\pm}=\pm\sqrt{\frac{a}{b}}$ and a quadratic maximum at $x_{0}=0$ of relative height $E_{B}=\frac{a^{2}}{4b}$. The
Hamiltonian of the quartic double well is then given by
\be
H_{S}(p_x,x)=\frac{p_x^2}{2m_x}+V_{S}(x).
\label{eq:dw}
\ee
Its classical dynamics consists of harmonic oscillations with the frequency $\omega=\sqrt{\frac{2a}{m_x}}$ close to the minima at $x_{\pm}$. These oscillations become increasingly
anharmonic as the energy rises towards the top of the barrier. The phase space portrait of the dynamics contains the prototypical separatrix, shaped like the number eight,
as well as a hyperbolic fixed point at $p_x=0,x=0$ on that separatrix and two elliptic fixed points at $p_x=0,x=x_{\pm}$ \cite{Reichl}. If the available energy is higher than the barrier,
in the case of the NH$_3$ molecule, the corresponding motion is referred to as umbrella motion.

\begin{figure}
\centering
\includegraphics[width=0.7\columnwidth,angle=0]{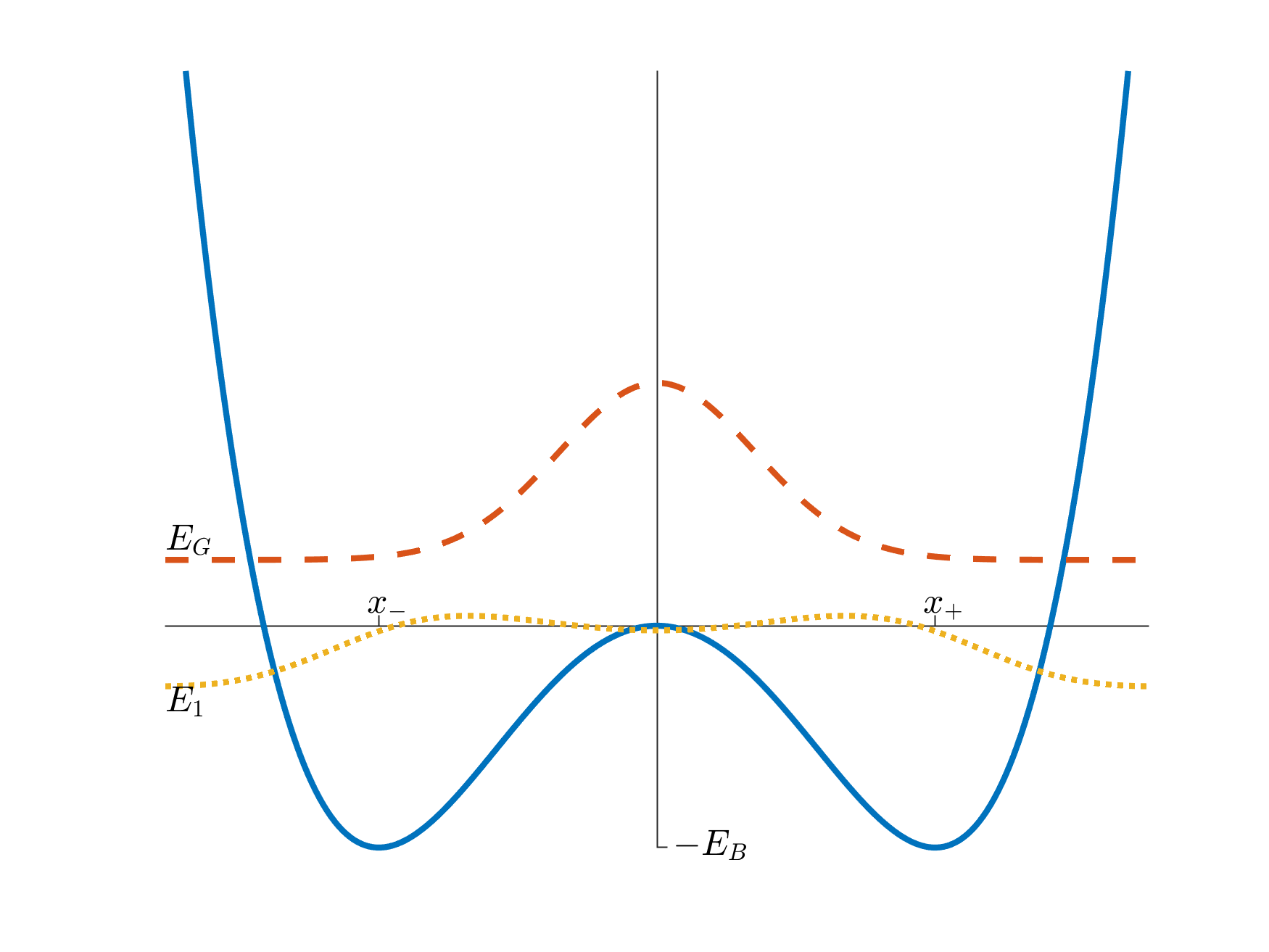}
\caption{Quartic double well potential with parameters $a=2$, $b=1$ (solid blue line), leading to $x_{\pm}=\pm\sqrt{2}$. Initial Gaussian sitting at the top of the barrier with initial 
energy $E_G=0.3$ (dashed orange line) and ground state of the quartic double well with ground state energy $E_{1}=-0.30$ (dotted yellow line).}
\label{fig:DW}
\end{figure}

For an understanding of the corresponding quantum dynamics, we first calculate the energy spectrum of the quartic double well potential, using a finite 
difference representation of the Laplacian to solve the time-independent Schr\"odinger equation (TISE)
\be
\hat{H}_{S}\phi_n(x)\equiv\left[-\frac{\hbar^2}{2m_x}\Delta+V_S(x)\right]\phi_n(x)=E_n\phi_n(x).
\ee 
The first few eigenvalues for the parameters $m_x=1$ and $a=2,b=1$ and in units, where $\hbar=1$ are gathered in Table \ref{tab:ev}.
\begin{table}
\begin{tabular}{c|c|c|c|c|c} 
&$n=1$&$n=2$ & $n=3$ & $n=4$&$n=5$ 
\\
\hline
$E_n$&-0.300&0.046&1.23&2.46&3.94
\\
\hline
$|c_n|^2$&0.654&0&0.323&0&0.0225
\end{tabular}
\caption{Eigenvalues $E_n$ and squared overlap $|c_n|^2:=|\langle \phi_n|\Psi(0)\rangle|^2$ of eigenstates with the initial Gaussian of Eq.\ (\ref{eq:DWini}) for $\gamma=2$ of the bare quartic double well with $a=2$, $b=1$.
The grid extension for the finite difference calculation was $x\in[-4,4]$ and 256 grid points were sufficient to achieve convergence to within the number of digits given.}
\label{tab:ev}
\end{table}
In the numerical work to be shown later, we will use a Gaussian of the form
\be
\Psi(x,0)=\left(\frac{\gamma_{x}}{\pi}\right)^{1/4}{\rm e}^{-\frac{\gamma_{x}}{2}x^2}
\label{eq:DWini}
\ee
that is located at the top of the barrier as the initial state. Together with the potential and its first eigenstate it is shown 
in Figure\ \ref{fig:DW}. The base lines for the two wavefunctions are at their corresponding energy expectation values,
which in the case of the Gaussian is $E_G=\gamma_{x}/4-1/(2\gamma_{x})+3/(16\gamma_{x}^2)$, leading to $E_G=0.3$ for a width parameter of
$\gamma_{x}=2$.
Although both, initial position as well as momentum expectations are zero, due to its finite width, there
is a finite energy content in the wavepacket.
This initial state is the motivation for the naming ``toppling pencil''; a pencil balanced tip down on a flat surface, prone to fall
over \cite{DPM20}.

For symmetry reasons, it is obvious that the symmetric initial Gaussian does have zero overlap with the eigenfunctions of odd parity (see Table \ref{tab:ev}).
 Like the odd ones, also almost all higher even eigenstates (from the fifth eigenstate on)
are not taking part in the dynamics
The dynamics of the Gaussian in the bare potential will then be an oscillation with the frequency corresponding to the difference of the first
and the third eigenvalue. We stress that this is not the usual tunneling scenario, where a Gaussian
is sitting in one of the two wells initially and is then moving to the other well and back with a (usually very small) frequency
given by the difference $E_2-E_1$ of the two lowest eigenvalues. Here we focus on the symmetric initial condition, however, and have chosen
the potential parameters such that just the  eigenstates with eigenvalues $E_1$ and $E_3$ are appreciably populated.

\subsection{Coupling to a Finite Heat Bath}

Coupling a double well to a harmonic oscillator heat bath with infinitely many degrees of freedom with continuous spectral density will lead to decoherence and dissipation in the dynamics.
In the case that only the two lowest states of the bistable system play a role, the tunneling rate will be severely influenced by the system bath coupling \cite{BM82}.
A lot of work has been done on that so-called spin-boson model in the 80s of the last century, as documented by the impressive review by Leggett et al \cite{Letal87}. More recently, this model has been studied deeply
by different numerical methods, with a focus on correctly mimicking infinite baths by either a discretization in frequency of the harmonic oscillator spectral density 
or by a correct description of the bath correlation function in the time domain \cite{jcp19-02}. Furthermore, also the influence of a sinusoidal driving on the 
tunneling effect in a two-level system has been studied, yielding surprising localization effects \cite{epl922,GH98}.

Here we are not restricting ourselves to the spin-boson case but will consider the total Hamiltonian to be that of the full bistable system coupled to 
an environment with a large but finite number of degrees of freedom, given by
\be
\label{eq:ham}
H({\bf R},{\bf r})=H_S({\bf R})+H_E({\bf r})+H_{SE}({\bf R},{\bf r})
\ee
where ${\bf R}=(p_x,x)$ denotes the phase-space vector of the central system of interest, whereas ${\bf r}=(p_1,p_2,...,p_f,y_1,y_2,...,y_{f})$ denotes the $2f$-dimensional phase space
vector of all the environmental degrees of freedom. The dynamics of the bare central system  of interest (index S) is governed by the Hamiltonian of the quartic
double well given in Eq.\ (\ref{eq:dw}). The environment (index E) consists of $f$ harmonic oscillators, whose Hamiltonian is given by
\be
H_E({\bf r})=\sum_{n=1}^{f}\left(\frac{p_n^2}{2m}+\frac{m\omega_n^2}{2}y_n^2\right).
\ee
The choice of a discrete set of frequencies $\omega_{n}$, $n=1,...,f$ will be discussed below in the numerical results section. 
As each oscillator should exert a force on the  system, hence their interaction can be modeled as the position-position coupling
\be
H_{SE}({\bf R},{\bf r})=-x\sum_{n=1}^{f}g_{n}y_n
\label{eq:hamint}
\ee
with coupling constant $g_n=\frac{g}{\sqrt{f}}$, which is renormalized by $\sqrt{f}$ in order to make the results for different number of oscillators comparable. Using linear response theory,
this scaling has been derived in \cite{MaAg11}. 

The bilinear coupling does not break the invariance of the total Hamiltonian under a parity transformation (spatial reflection) $P: ({\bf R},{\bf r})\rightarrow(-{\bf R},-{\bf r})$. However, it drives 
the two bistable minima apart
from $x_{\pm}=\pm\sqrt{\frac{a}{b}}$ to $x_{\pm}=\pm\sqrt{\frac{1}{b}(a+\sum_{n=1}^{f}\frac{g_{n}^{2}}{m\omega_{n}^{2}})}$, an effect which is not intended by 
coupling to the environment \cite{DPM20}. This driving apart can, however, be compensated by including the so-called counter term proportional to the square of the system coordinate 
in the potential of the total Hamiltonian (system plus bath) to complete the squares with respect to the dependence on the oscillator coordinates. One thus replaces the total potential
by \cite{In02,DPM20}
\begin{align}
V_{S}+V_E+H_{SE}+V_C&=V_{S}(x)+\sum_{n=1}^{f}\frac{m\omega_n^2}{2}y_n^2-x\sum_{n=1}^{f}g_{n}y_n+x^{2}\sum_{n=1}^{f}\frac{g_{n}^{2}}{2m\omega_{n}^{2}}
\nonumber
\\
&=V_{S}(x)+\sum_{n=1}^{f}\frac{m\omega_n^2}{2}\left(y_n-\frac{g_n}{m\omega_{n}^{2}}x\right)^{2}.
\end{align}
In the section on the numerical results, we will display the time evolution of the counter term $V_C$.
Even more importantly, we will start the bath in its ground state (in the thermodynamic limit $f\to \infty$ this
would correspond to zero temperature) and will focus on the evolution of the system dynamics away from
the excited state on top of the barrier. To this end we will need a numerical method that allows us to treat a multitude of
degrees of freedom quantum mechanically. The method of our choice is the CCS method to be reviewed in the following.

\section{Coupled Coherent States Method}

To tackle the quantum dynamics under the many body Hamiltonian from the
previous section, grid based methods are running into an an exponential wall and 
we will use a variational approach, based on time-evolving coherent states, that has been
introduced by Shalashilin and Child \cite{SC00,ShBu08}, and was recently reviewed in \cite{irpc21}.
In the following we will briefly recapitulate the so-called coupled coherent state  
Ansatz and its working equations, extending the notation of \cite{irpc21} towards many degrees of freedom.

For a system of $(f+1)$ degrees of freedom as in the previous section, an Ansatz for the solution of the time-dependent Schr\"odinger equation (TDSE) is given in terms of 
multi-mode coherent states (CS) of multiplicity $M$ by
\be
\ket{\subps{\Psi}{CS}{M}(t)}=\sum_{l=1}^M a_l(t)\ket{\bm{z}_l(t)}, \label{eq:D_ansatz}
\ee
with time-dependent complex coefficients $a_k(t)$ and time-dependent $(f+1)$-dimensional complex displacement vectors 
\be
{\bm z}_l(t)=\frac{{\boldsymbol \gamma}^{1/2}{\bm q}_l(t)+{\rm i}{\boldsymbol \gamma}^{-1/2}{\bm p}_l(t)}{\sqrt{2}},
\ee
with ${\bm q}=(x,y_1,\dots,y_f)$ and ${\bm p}=(p_x,p_1,\dots,p_f)$ and diagonal matrix ${\boldsymbol \gamma}$ with entries
$\gamma_j=m_j\omega_j,\quad j=0,\dots,f$, where $\gamma_0=\gamma_{x}$, $m_0=m_x$ and $m_1,\dots,m_f=m$ and we have set
$\hbar=1$.

The $(f+1)$-mode CS are given by an $(f+1)$-fold tensor product
\be
\ket{{\bm z}_l} = \bigotimes_{j=0}^{f}\ket{z_{lj}}
\ee
of normalized one-dimensional CS
\be
\ket{z_{lj}}=\exp\left[-\frac 12|z_{lj}|^2\right]\exp\left[z_{lj}\hat a_j^\dagger\right]\ket {0_j},  
\ee
where $\hat a_j^\dagger$ is the creation operator acting on the ground state
of a suitably chosen $j$-th harmonic oscillator and the CS form an over-complete and nonorthogonal basis set \cite{BBGK71}
and are Gaussian wavefunctions in position space.

To make progress, the Hamiltonian in (\ref{eq:ham}) is to be expressed in terms of 
the creation and annihilation operators of the harmonic oscillator underlying 
the CS. In all that follows, we will use the normally ordered Hamiltonian, where all appearances of $\hat{a}_j^\dagger$ precede those of $\hat{a}_j$.
Whereas for the bath part of the Hamiltonian, which is harmonic, the task of finding the normally ordered Hamiltonian is trivial, 
for the quartic bistable system of interest, corresponding to index $j=0$,
we give the derivation of the corresponding expression in some detail in Appendix \ref{app:noeom}.

The time-evolution of the coefficients and the displacements is now governed by the Dirac-Frenkel 
variational principle \cite{Di30,Fren34}
\be
\bra{\delta\subps{\Psi}{CS}{M}}{\rm i}\partial_t-\hat{H}\ket{\subps{\Psi}{CS}{M}}=0,\label{eq:dirac_frenkel}
\ee
and we  have given the fully variational equations of motion in \cite{prb20}.
In the case that the Ansatz (\ref{eq:D_ansatz}) is restricted to a single
term, i.e., $M=1$, these equations reduce to
\begin{align}
{\rm i}\dot{a}&=a\left[H_{\rm ord}({\bm z}^\ast,{\bm z})
-\frac{\rm i}{2}(\dot{{\bm z}}\cdot{\bm z}^\ast-{\bm z}\cdot\dot {\bm z}^\ast)\right],
\\
\label{eq:cl}
{\rm i}\dot{{\bm z}}&=\frac{\partial H_{\rm ord}}{\partial {\bm z}^\ast}.
\end{align}
The second of these equations are the complexified Hamilton's equations and they are given for the present case in Appendix \ref{app:noeom}.

In the CCS method one now reintroduces the multiplicity index and propagates all the coherent state parameters ${\bm z}_l(t)$ in the Ansatz (\ref{eq:D_ansatz}) according 
to the classical equations and keeps the fully variational equations of motion for the coefficients $a_l(t)$ \cite{ShBu08}, given by
\be
\label{eq:ccs}
{\rm i}\sum_{l=1}^M\langle {\bm z}_k(t)|{\bm z}_l(t)\rangle\dot{a}_l(t)=\sum_{l=1}^M{\tilde H}_{kl}(t) a_l(t),
\ee
with the time-dependent matrix elements (even in the case of an autonomous Hamiltonian)
\begin{align}
{\tilde H}_{kl}(t)&=\langle {\bm z}_k(t)|{\bm z}_l(t)\rangle
\Biggl[H_{\rm ord}({\bm z}_k^\ast,{\bm z}_l)-
\nonumber\\
&\frac{1}{2}\left({\bm z}_l(t)\cdot
\frac{\partial H_{\rm ord}({\bm z}_l^\ast,{\bm z}_l)}{\partial {\bm z}_l}-\frac{\partial H_{\rm ord}({\bm z}_l^\ast,{\bm z}_l)}{\partial {\bm z}_l^\ast}\cdot{\bm z}_l^\ast(t)\right)
-{\bm z}_k^\ast(t)\cdot\frac{\partial H_{\rm ord}({\bm z}_l^\ast,{\bm z}_l)}{\partial {\bm z}_l^\ast}\Biggr],
\end{align}
where the elements of the overlap matrix of the multi-mode CS are given by
\begin{align}
\langle{{\bm z}_k}|{{\bm z}_{l}}\rangle&=\prod_{j=0}^{f}\exp\left[-\frac{1}{2}\left(|z_{kj}|^2+|z_{lj}|^2\right)+z_{kj}^\ast z_{lj}\right]\nonumber\\
&=\exp\left[-\frac 12(|{\bm z}_l|^2+|{\bm z}_{k}|^2)+{\bm z}_k^\ast\cdot{\bm z}_l\right]
\nonumber\\
&=\exp\left[-({\bm q}_l-{\bm q}_k)^{\rm T}\frac{{\boldsymbol \gamma}}{4}({\bm q}_l-{\bm q}_k)-({\bm p}_l-{\bm p}_k)^{\rm T}\frac{{\boldsymbol \gamma}^{-1}}{4}({\bm p}_l-{\bm p}_k)
+\frac{\rm i}{2}({\bm p}_l\cdot{\bm q}_k-{\bm q}_l\cdot{\bm p}_k)\right],
\label{eq:over}
\end{align}
which is the product of the corresponding single mode overlaps. We note that the Klauder phase convention
of the corresponding Gaussian wavepackets has been used \cite{irpc21}.

For the determination of the initial conditions of the trajectories, we are using the pancake sampling idea suggested by
Shalashilin and Child \cite{SC08}. It is the (random) sampling of the initial conditions in the extended (system plus bath) phase space
that is believed to help the CCS method cope with the exponential wall, that is usually encountered in grid-based approaches
to many-body quantum dynamics.

\section{Long-time dynamics of the coupled system}

In the following, we will present numerical results for the time-evolution of the composite system
using the method just described to solve the time-dependent Schr\"odinger equation. In addition, for up 
to a total of 4 DOF, i.e., $f=3$, we also corroborated our results by using the split-operator fast Fourier transform (FFT) technique for 
quantum propagation \cite{FFS82}\footnote{We note that the number of grid points in the $x$ direction needed for convergence
was just 32 (for the grid extension $x\in [-3.5,3.5]$) and thus much less than in the finite difference approach in \ref{ssec:DW}. Furthermore, the number of grid points we took for the
harmonic degrees varied between 128 for the low frequency oscillators and 64 for the high frequency ones. The time-step for propagation was $\Delta t=0.01$.}.
Our focus will be on the question if the coupling to the environmental degrees of freedom,
which are all starting in their ground states
\be
\Psi(y_n,0)=\left(\frac{\gamma_n}{\pi}\right)^{1/4}\exp\left\{-\frac{\gamma_n}{2}y_n^2\right\},
\label{eq:envini}
\ee
will eventually lead to a ``thermalization'' of the quartic degree towards its ground state
(which is depicted in Fig.\ \ref{fig:DW} as the dotted yellow line). We stress that previous treatments
of the double well dynamics using CCS \cite{SDC06} have focussed on the description of quantum tunneling, where the initial state
is made up of an equal weight superposition of the two lowest energy eigenstates, whereas herein, the initial
states consists mostly of state one and state three (see Table \ref{tab:ev}).

In the following, the potential parameters for the double well are $a=2$ and $b=1$ and the mass $m_x$ is set equal to unity.
The choice of frequencies of the environmental degrees of freedom will be detailed below. All masses
of the oscillators are taken to be equal and given by $m=0.1$. 

\subsection{Different numerical measures}

As a first measure of the possible deviation of the time-evolved wavefunction away from the initial state,
we use the autocorrelation function, defined in 1D as
\begin{align}
c(t)=\langle \Psi(0)|\Psi_{\rm CS}^M(t)\rangle&=\sum_{l=1}^Ma_l(t)\langle \Psi(0)|z_l(t)\rangle
\nonumber
\\
&=\sum_{k,l}a_k^\ast(0)\langle z_k(0)|z_l(t)\rangle a_l(t).
\label{eq:cf1D}
\end{align}
For a multitude of degrees of freedom, an analogous quantity could be defined by just replacing
the scalars $z_k,z_l$ by the corresponding vectors ${\bm z}_k,{\bm z}_l$, which  would, however,
not serve our purpose. Our goal is to find an autocorrelation measure, irrespective of the dynamics
of the environment. Therefore, we first define the probability density of the system degree
of freedom, by integrating the absolute value squared of the full wavefunction over all $N$ environmental
degrees of freedom
\be
\label{eq:den}
\rho_S(x,t)=\int{\rm d}y_1\dots{\rm d}y_f|\Psi_{\rm CS}^M(x,y_1,\dots,y_f,t)|^2
\ee
to arrive at the probability density of the quartic degree of freedom.
This then allows to calculate the quantity
\be
c_S(t)=\int{\rm d}x\Psi(x,0)\sqrt{\rho_S(x,t)},
\label{eq:acND}
\ee
which (if the initial state is real (and positive) as herein) has no phase anymore and is thus the analogue of the absolute
value of the correlation function of Eq.\ (\ref{eq:cf1D}). For the pure quartic (1D) case and the initial state we use,
the time evolution of the quantities defined Eq.\ (\ref{eq:cf1D}) and Eq.\ (\ref{eq:acND}) is similar but not idential. The oscillation period
is identical though.

An even more stringent measure to decide if the time evolved state is approaching the ground state
is the energy expectation value, defined by
\begin{align}
\langle E\rangle(t)&=\langle \Psi_{\rm CS}^M(t)|\hat{H}_{\rm ord}|\Psi_{\rm CS}^M(t)\rangle
\nonumber
\\
&=\sum_{k,l}a_k^\ast(t)H_{\rm ord}({\bm z}_k^\ast,{\bm z}_l)\langle {\bm z}_k(t)|{\bm z}_l(t)\rangle a_l(t).
\end{align}
The different terms in the total Hamiltonian can be disentangled and their respective contribution
to the total energy can be looked at separately. The conservation of the total energy will also serve as
a convergence check for the CCS method \cite{Haber12}. In passing, we note that the norm of the total wavefunction
was well conserved in all the numerical calculations that we present. This is in contrast to the semiclassical
Herman-Kluk case, where often a normalization of the results has to be performed \cite{TS99}. This is not necessary for CCS.

\subsection{Numerical results}

\begin{figure}
\centering
\includegraphics[width=0.7\columnwidth,angle=0]{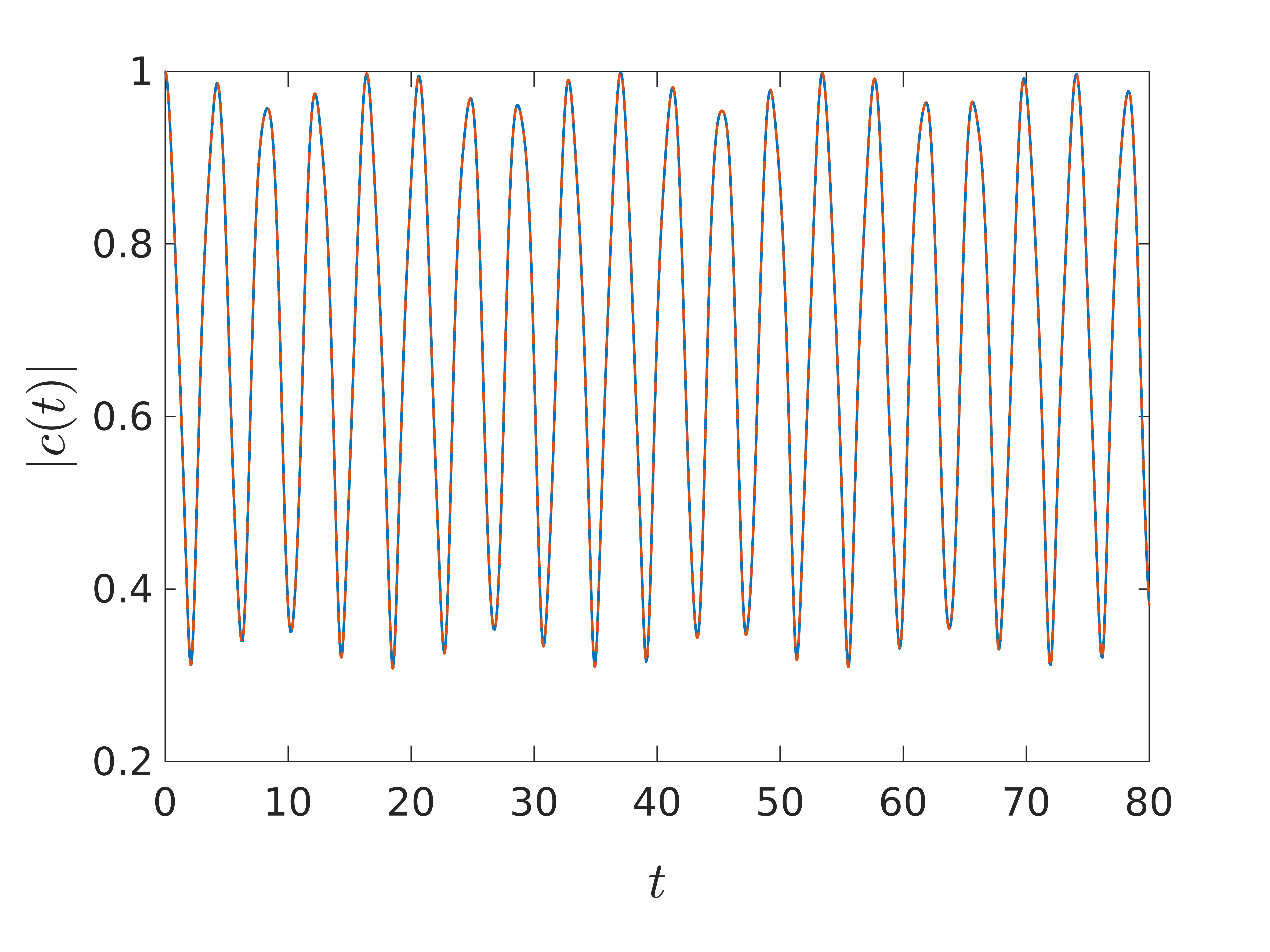}
\caption{Absolute value of the autocorrelation function for the 1D system (double well only)
from a CCS calculation (solid blue line) and from a split-operator FFT calculation (dashed red line).
The two curves coincide to within line thickness.}
\label{fig:ac1D}
\end{figure}

First, we consider the autocorrelation of the initial state at the top of the barrier in the uncoupled 1D case in Fig.\ \ref{fig:ac1D}. 
For the CCS calculations, a multiplicity of $M=299$ was enough to  converge the result to the converged split operator FFT result.
Because the initial state mainly contains only two eigenstates (see Table \ref{tab:ev}), the local spectrum  \cite{Gross3}, i.e., the Fourier transform
of the autocorrelation contains just two major peaks. From the figure,
it can be seen that the absolute value of the autocorrelation correspondingly oscillates back and forth between unity and around 
0.3 with a period of around $T=2\pi/(E_3-E_1)\approx 4$ in
the dimensionless units used. The initial state will thus be revisited frequently in the uncoupled case.

The initial state that will be used in the propagation of the coupled system is the (direct) product of the Gaussian
on top of the barrier of the double well, given in Eq.(\ref{eq:DWini}), times the ground state
Gaussians of Eq.\ (\ref{eq:envini}). Before showing the alteration of the results by the coupling of the double well to several oscillators, 
we have to elaborate on the choice of frequencies of those oscillators, however. This choice will be crucial for the energy 
flow between the double well and the environment. We again follow the work of DPM \cite{DPM20}
and choose the frequencies from the (normalized) density
\be
\rho_\omega(\omega)=\frac{1}{\omega_{\rm co}}\exp\left\{-\frac{\omega}{\omega_{\rm co}}\right\},
\ee
with a parameter $\omega_{\rm co}$ to be fixed below,  according to
\be
\int_0^{\omega_k}{\rm d}\omega \rho_f(\omega)\stackrel{!}{=}\frac{k}{f_{\rm co}}.
\ee
This leads to the explicit expression
\be
\omega_k=-\omega_{\rm co}\ln\left(1-\frac{k}{f_{\rm co}}\right)\,\qquad k=1,....f
\label{eq:dis}
\ee
for the frequencies and we here choose the second parameter $f_{\rm co}>f$, such that extremely high frequencies which would
not exchange energy with the system (not shown) are not considered. Other frequency distributions have been used in
\cite{jcp19-02} as well as in \cite{cp102}, while the present one has been found favorable also in multi configuration time-dependent Hartree (MCTDH)
calculations \cite{WT08}. Now one could choose the coupling strength between system and environment according 
to a specific (continuous) spectral distribution, which is usually taken as Ohmic or sub- or super-Ohmic. Here, however, 
we again adhere to DPM and take equal coupling strengths $g=0.1$ for all oscillators, just suitably normalized
by the total number of environmental degrees of freedom (see remark after Eq.\ (\ref{eq:hamint})) to make the results
for different values of $f$ comparable. In Table \ref{tab:dis}, we give the parameters that were used in
Eq.\ (\ref{eq:dis}) for the calculation of the discretized frequencies for different values of $f$.
In the following several different quantities will be looked at for increasing numbers of environmental degerees of freedom.
\begin{table}
\begin{tabular}{c|c|c|c|c} 
$f$&$2$&$3$&$4$&$5$ 
\\
\hline
$\omega_{\rm co}$&4&4&4&4
\\
\hline
$f_{\rm co}$&10&12&14&16
\end{tabular}
\caption{Parameters needed for the  calculation of  the discretized bath frequencies according to Eq.\ (\ref{eq:dis})
for different values of $f$.}
\label{tab:dis}
\end{table}

\begin{figure}
\centering
\includegraphics[width=0.7\columnwidth,angle=0]{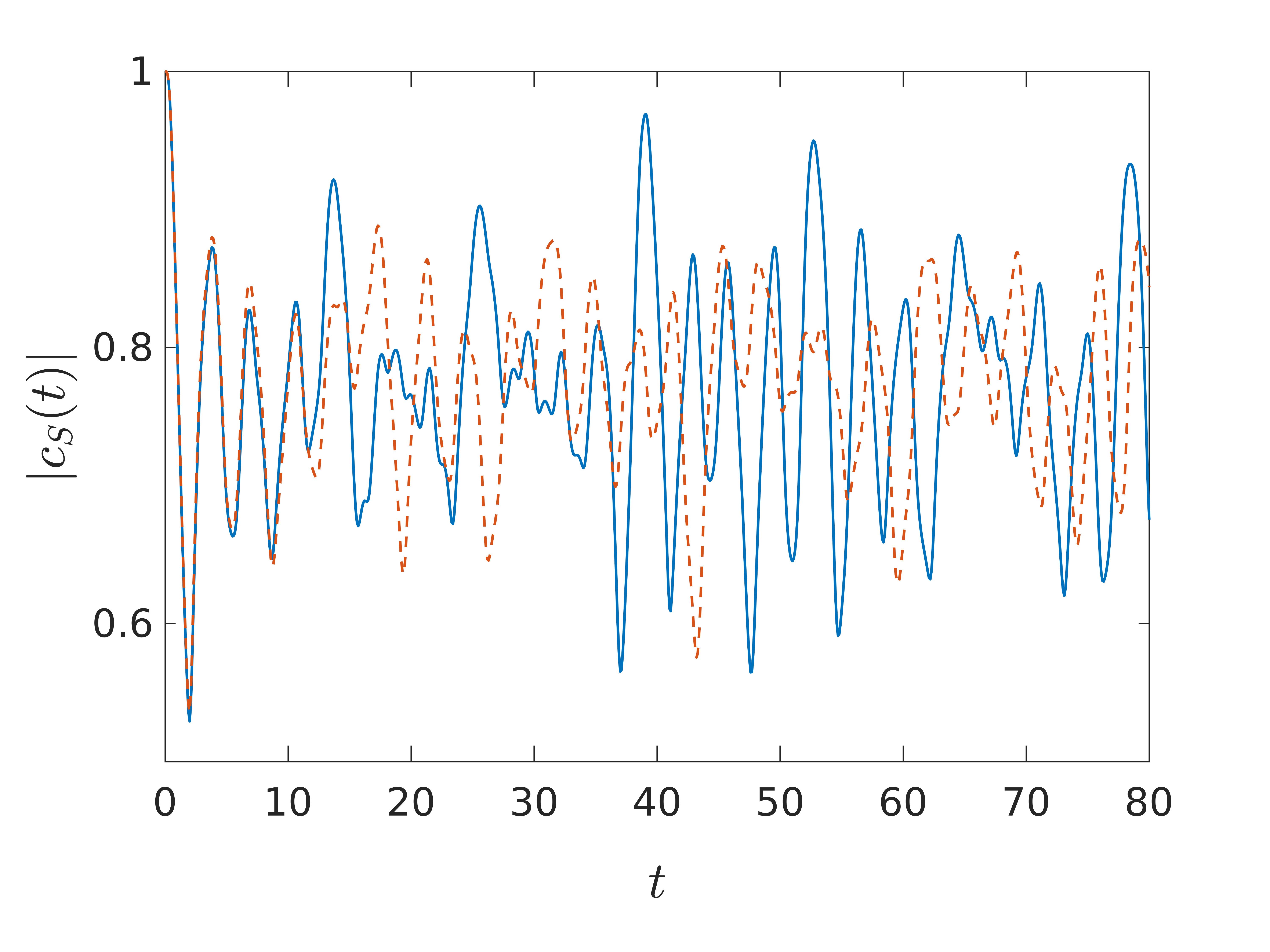}
\caption{Absolute value of the autocorrelation-like quantity defined in Eq.\ (\ref{eq:acND})
from a CCS calculation for different numbers of environmental oscillators: $f=2$ (solid blue) and $f=3$ (dashed red) }
\label{fig:acND}
\end{figure}

To start with, for the system's ``autocorrelation'' $c_S(t)$ defined in Eq.\ (\ref{eq:acND}) , we found the results displayed in Fig.\ \ref{fig:acND}.
For the 3D results ($f=2$) we used 799 trajectories, whereas for the 4D case ($f=3$) we used 2999.
From Fig.\ \ref{fig:acND} and by comparison to the 1D case displayed in Fig.\ \ref{fig:ac1D},
it can be seen that by the coupling to the environmental degrees of freedom, the oscillation frequency is only marginally increased 
(as to be expected by comparison to Rabi oscillations in a two level system) 
but the oscillation amplitude becomes decisively smaller. Furthermore, a damping of the oscillations for long times 
can be observed, which becomes the more prominent, the higher is the number of environmental degrees of freedom.
If the quartic subsystem evolves towards the ground state for long times, the expected long-time asymptotic value of the quantity we calculated is $c_S(\infty)=|c_1|\approx 0.8$
(taking the square root of $|c_1|^2$ from Table \ref{tab:ev}). 
This value is close to the asymptotic average value of the results displayed in Fig.\ \ref{fig:acND}.

\begin{figure}
\centering
\includegraphics[width=0.49\columnwidth,angle=0]{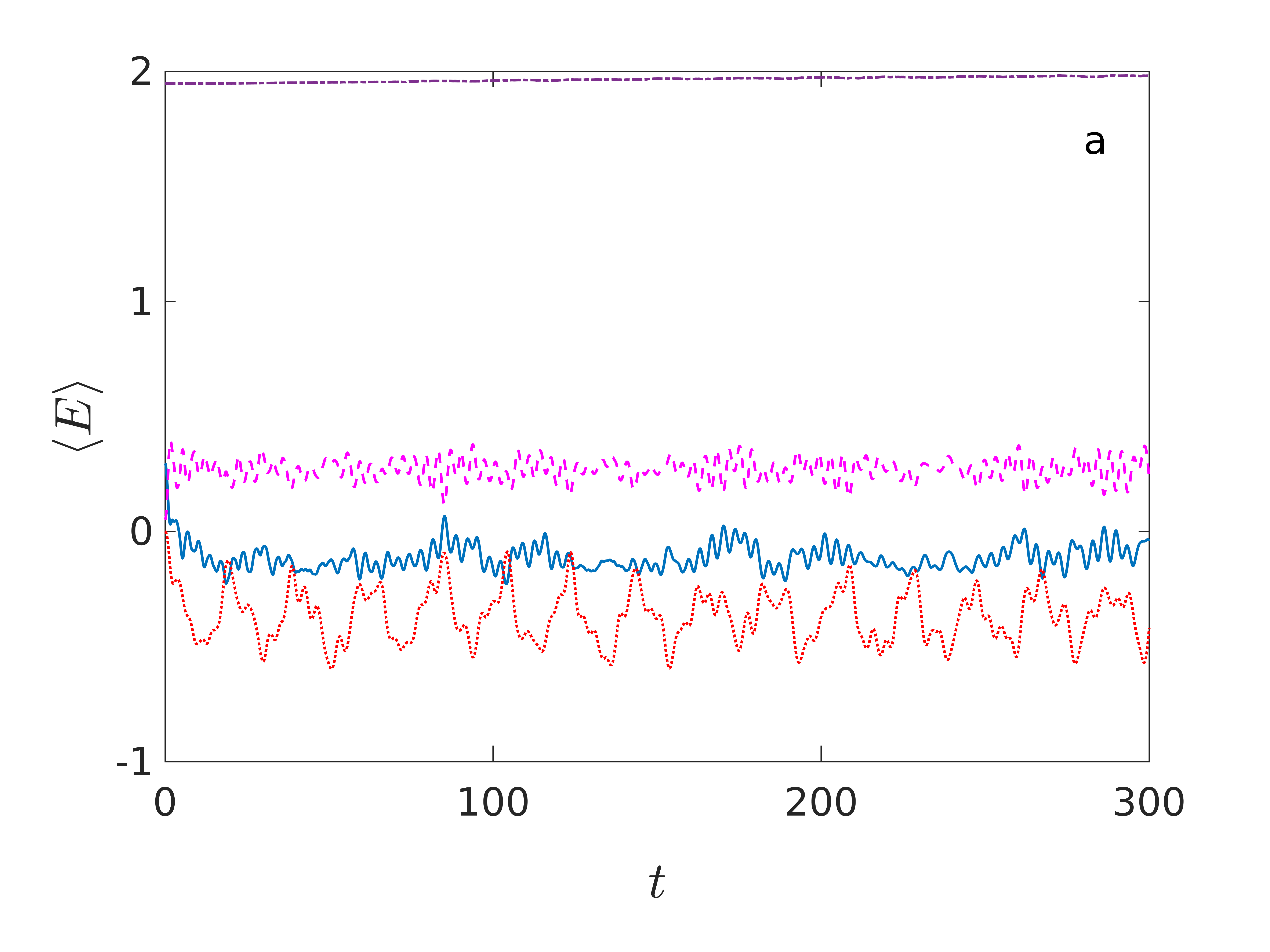}
\includegraphics[width=0.49\columnwidth,angle=0]{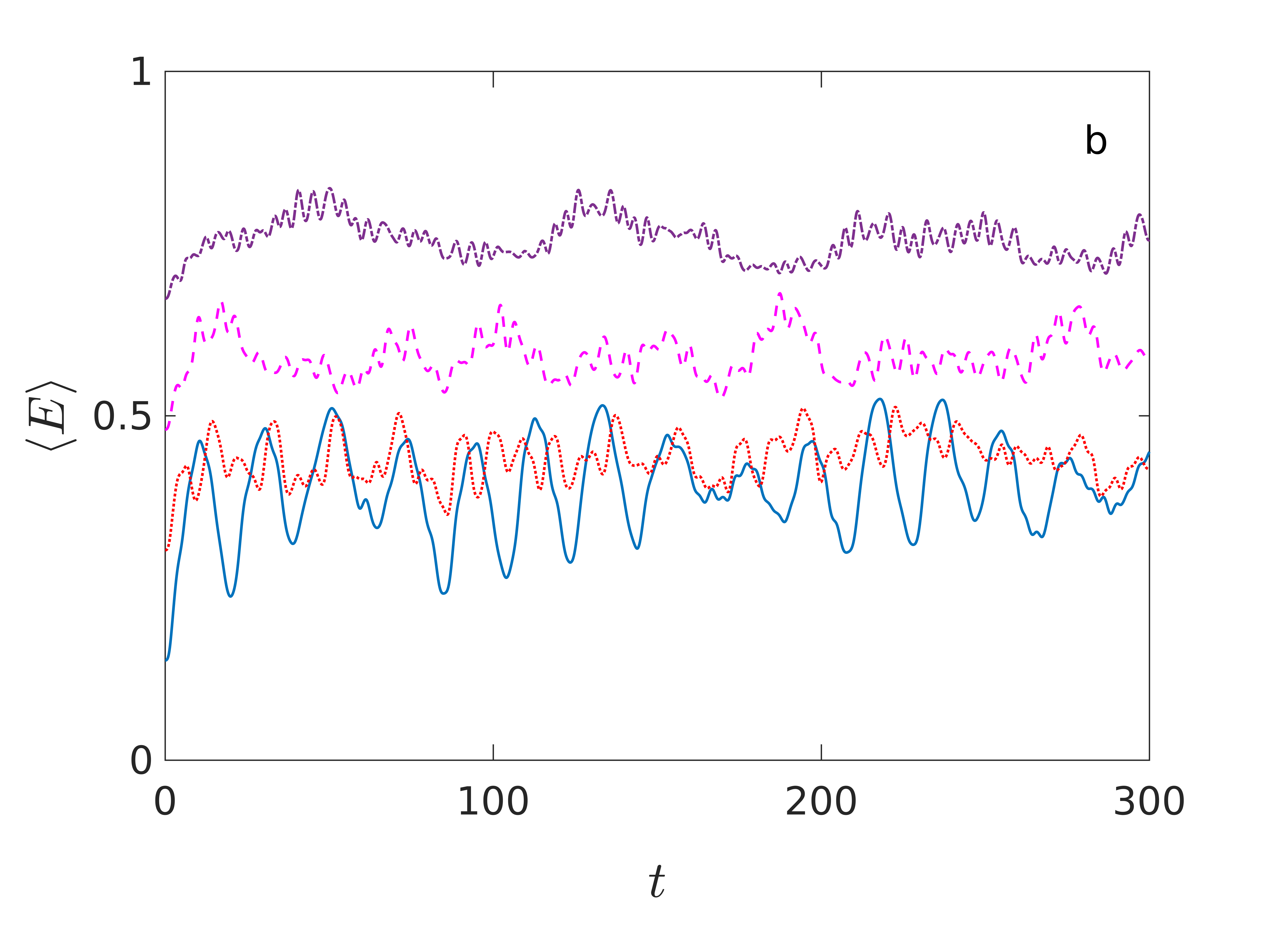}
\caption{Different energy expectation values for the case $f=4$: (a) Total energy (dashed-dotted), counter term (dashed), double well (solid), coupling term (dotted),
(b) kinetic plus potential energy of the different harmonic degrees of freedom: $\omega_1=0.29$ (solid), $\omega_2=0.61$ (dotted), $\omega_3=0.96$ (dashed) $\omega_4=1.34$ (dashed-dotted)).} 
\label{fig:en5D}
\end{figure}

The most important measure to decide about the propagated density's possible evolution towards the ground state
is the energy expectation value. In Fig.\ \ref{fig:en5D} the energy expectations for all the 5 degrees of
freedom in the case of $f=4$ as well as the coupling and the counter term are displayed, whereby panel (a) contains the total energy, the double well energy
as well as the coupling and the counter term and panel (b) contains all the different oscillator energies. The total 
energy is conserved very well, although a small tendency towards an energy drift is visible (panel (a)). For the presented
results we have used a multiplicity of $M=5999$ and we did not increase this number because the convergence 
is becoming exceedingly slow with the number of trajectories and the presented calculation took already several days on a modern computer cluster using several cores.
As displayed in panel (b), the different oscillators clearly show an increase in their energy, away from their ground state value and the $\omega_1$ oscillator even overtakes
the $\omega_2$ oscillator in terms of energy at certain times. We stress that the oscillators' energies never fall below their ground state energies as it should be \cite{cp18}. 
Furthermore, the oscillator with the highest frequency is still showing appreciable variations in its energy. If even higher frequencies would
have been chosen, the energy transfer would start to diminish, however (not shown). We stress that the environment, by consisting of a finite number of degrees of freedom,
does heat up, in contrast to the case of infinitely many environmental oscillators described by a continuous spectral density \cite{Tan20}.
As shown in panel (a), the counter term shows high frequency oscillations with a period similar to the
total double well energy and the interaction energy is large and negative. Most importantly, the total double well energy
shows a clear tendency to decrease below its initial value of 0.3. In addition, for large times, the amplitude of oscillation
of the double well energy around its average value is rather small. 

\begin{figure}
\centering
\includegraphics[width=0.7\columnwidth,angle=0]{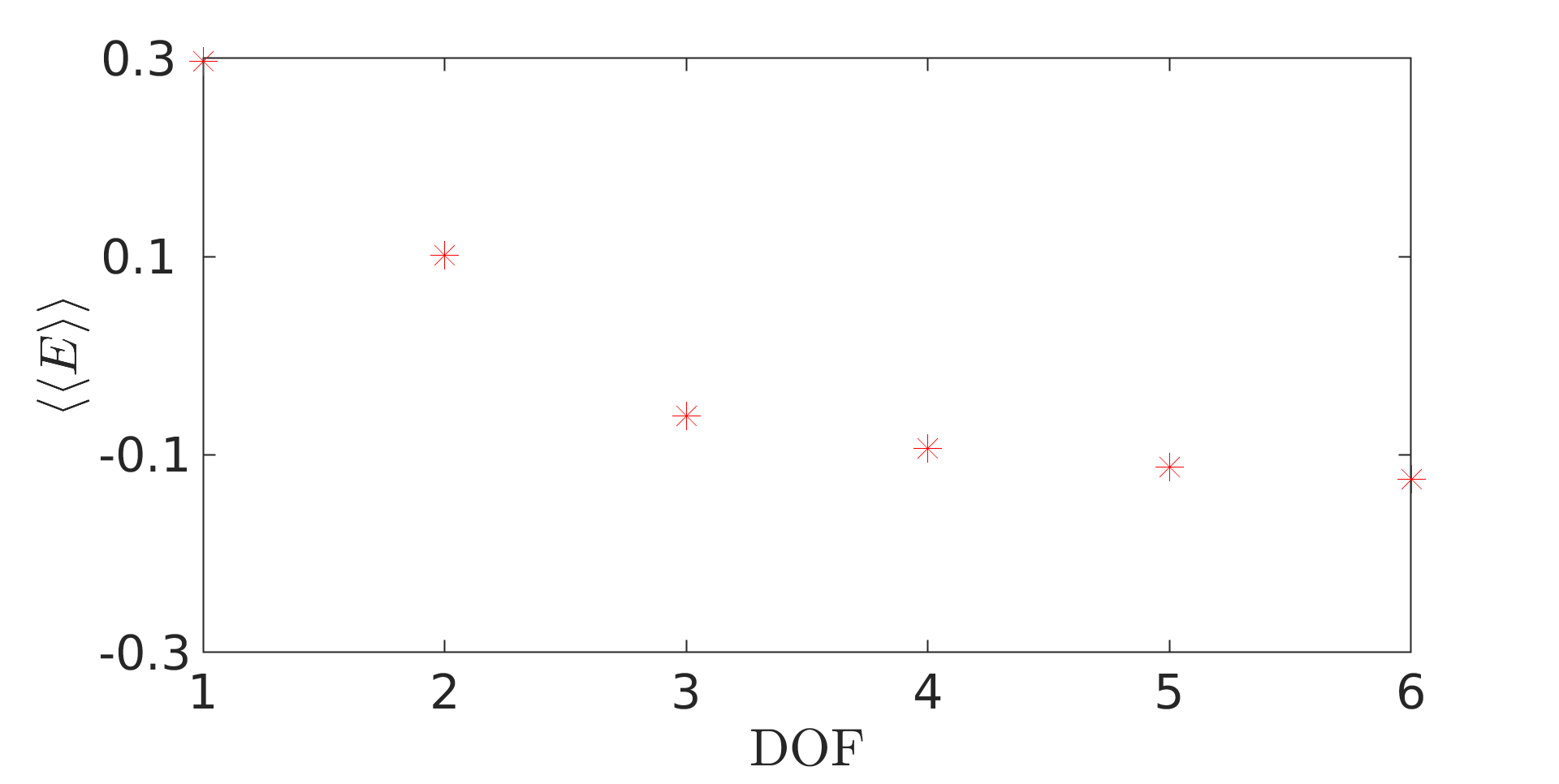}
\caption{Long-time average energy of the double well as a function of the total number of degrees of freedom (DOF).}
\label{fig:aven}
\end{figure}

This last finding led us to investigate the long-time average value of the double well
energy expectation
\be
\langle\langle E\rangle\rangle=\frac{1}{T_{\rm tot}}\int_0^{T_{\rm tot}}{\rm d}t\langle E\rangle(t)
\ee
as a function of the number of environmental oscillators for the relatively long total time of $T_{\rm tot}=300$.
As shown in Fig.\ \ref{fig:aven}, the average energy of the double well oscillator indeed decreases as a function
of the number of environmental oscillators, indicating a clear trend towards the ground state energy.
In the supplementary material, we provide a video of the time evolution of the probability density of the double well degree of freedom, defined in Eq.\ (\ref{eq:den}), for $0<t<80$ \footnote{In the supplementary video (link to be provided by publisher) 
the evolution of the 1D probability density $\rho_S(x,t)$ from the coupled dynamics with $f=3$ harmonic oscillators is displayed on top of the ground state density (static blue line) of the unperturbed double well potential
in the interval $0<t<80$}. This video shows
that for long times, the time-evolved density, due to the coupling (in the presented case to 3 harmonic degrees of freedom) approaches the (static) ground state density (also displayed in the movie) to a substantial degree.

Finally, it is worthwhile to mention that we have tried different values for the coupling strength $g$ and found (not shown) that larger values
lead to too much initial energy from the counter term (the coupling itself has zero expectation value at $t=0$) and thus eventually in the system, whereas smaller values of $g$, due to the small 
coupling decrease the flow away
from the quartic into the harmonic degrees of freedom. So the value $g=0.1$ that we used was close to optimal
if the cooling of the quartic degree of freedom is to be achieved.

\section{Conclusions and Outlook}

By focusing on the toppling pencil model, i.e., an excited initial state on top of the
barrier of a symmetric quartic double well, we have investigated if the coupling of
the central quartic double well to a finite, environmental bath of harmonic oscillators
in their ground states will let the central system evolve towards its ground state.
This amounts to the thermalization, i.e., a cooling down to the bath ``temperature''
(strictly only defined in the thermodynamic limit) of the central system.

By solving the time-dependent Schr\"odinger equation using the CCS methodology (and also split-operator FFT for 
small numbers of degrees of freedom), we could show that, indeed, the coupling eventually excites those environmental oscillators
and for the relatively long times investigated, there is no appreciable back flow of energy to the system of interest, such that the
central degree of freedom looses an appreciable amount of energy, monitored by its long-time average,
which we found to be a monotonously decaying function of the number of environmental degrees of freedom. 
For the largest number $f=5$ of oscillators that we investigated, the long-time average of the double well energy 
decreases from its uncoupled value of 0.3 to $-0.125$. This tendency of driving the double well towards lower energy via the environmental coupling
was corroborated by an autocorrelation function-like measure,
which showed a long-time behaviour in accord with the estimate assuming a total transition to the ground state of the quartic degree of freedom.
In the light of these results it is worthwhile to note that the
variational approach to solving the TDSE is based on a Lagrangian, see, e.g., \cite{ShBu08} and the wavefunction
parameters can be viewed as classical generalized coordinates. There are usually many more of them than in the
pure classical approach where there is only a pair of position and momentum variables per degree of freedom.

We have thus added a further example to the list of continuous variable systems of interest with (in principle) infinite dimensional 
Hilbert space that are coupled to a finite bath and show signs of energetic equilibration. Up to now, the focus in the literature was mainly on
spin-chain systems \cite{JeSh85,GMM09,HFT15}. The bath, by consisting of a finite number of degrees of freedom, also has a finite
heat capacity, in contrast to treatments in terms of reduced density matrices that hinge on
continuous spectral densities of the oscillator bath \cite{Wei08,Tan20}. By (hypothetically) going to the limit 
of very large $f$, due to the scaling of the coupling by the square root of $f$, the coupling to the individual oscillators
will be diminished and thus also the energy flow, such that the oscillators will stay close to their
initial state, which was here one of zero temperature (all oscillators in their ground states). Due to limited
computer resources the explicit fully quantum treatment of all degrees of freedom will be difficult in these cases though.
Other possible methods to try in the future could be MCTDH \cite{BJWM00}, matrix product state \cite{BoGe20} or different
types of semiclassical treatments \cite{jcp06,irpc21}.

Furthermore, so far the dynamics studied does not break the symmetry of the initial state of the double well, because the
Hamiltonian as well as the initial state of the bath are symmetric under the parity transformation. In the classical mechanics study by
Dittrich and Pena Mart{\'i}nez \cite{DPM20}, it has been argued that small asymmetries in the initial condition of the
bath lead to a preference of the bistable system to end up in one of the wells. It will be a worthwhile endeavor
to study such asymmetric initial conditions also in a full quantum time-evolution.

{\bf Acknowledgments:}
The authors would like to thank Profs.\ Marcus Bonanca and Thomas Dittrich for enlightening discussion 
as well as the International Max Planck Research School for Many Particle Systems in Structured Environments
for its support. FG would like to thank the Deutsche Forschungsgemeinschaft for financial support under grant GR 1210/8-1.

\begin{appendix}

\section{Normally ordered Hamiltonian and classical equations of motion}
\label{app:noeom}
The Hamilton operator for the quartic double well potential is given by
\begin{align}
\hat{H}_{S}&=\hat{T}_{S}+\hat{V}_{S}=\frac{\hat{p}_{x}^{2}}{2m_{x}}-\frac{a}{2}\hat{x}^{2}+\frac{b}{4}\hat{x}^{4}
\end{align}
To make progress, the position and momentum operators are expressed via creation and annihilation operators, via
\be
\hat{x}=\frac{1}{\sqrt{2\gamma_{x}}}(\hat{a}_{x}^\dagger+\hat{a}_{x}),\qquad
\hat{p}_{x}={\rm i}\hbar\sqrt{\frac{\gamma_{x}}{2}}(\hat{a}_{x}^\dagger-\hat{a}_{x}).
\ee
In the following we set $\hbar=1$. The subscripts to the creation and annihilation operators denote the respective degree of freedom. 
Using the fundamental commutation relation 
\be
[\hat a_x,\hat a_x^\dagger]=\hat 1,
\ee
the normal ordered form of the kinetic energy operator is found to be
\be
\hat T_{S}(\hat a_{x}^\dagger,\hat a_{x})=\frac{\hat{p}_{x}^2}{2m_{x}}=-\frac{\gamma_{x}}{4m_x}(\hat a_{x}^\dagger-\hat a_{x})^2=-\frac{\gamma_{x}}{4m_x}\left[(\hat{a}_{x}^\dagger)^2-2\hat a_{x}^\dagger\hat a_{x}-1+\hat a_{x}^2\right]
\ee
and thus
\be
\label{eq:CSkin}
T_{S}(z_{kx}^\ast,z_{lx})=-\frac{\gamma_{x}}{4m_{x}}\left[\left(z_{kx}^\ast-z_{lx}\right)^{2}-1\right]
\ee
follows for the CS matrix elements of the normally ordered kinetic energy operator.

Now, for the potential energy operator $\hat{V}_{S}$ of the system, we have
\begin{align}
\hat{x}^{2}&=\frac{1}{2\gamma_{x}}(\hat{a}_{x}^\dagger+\hat{a}_{x})(\hat{a}_{x}^\dagger+\hat{a}_{x})\nonumber\\
&=\frac{1}{2\gamma_{x}}\left[(\hat{a}_{x}^\dagger)^{2}+2\hat{a}_{x}^\dagger\hat{a}_{x}+1+\hat{a}_{x}^{2}\right]
\end{align}
and
\begin{align}
\label{eq:fourpower}
\hat{x}^{4}=\hat{x}^{2}\hat{x}^{2}&=\frac{1}{4\gamma_{x}^{2}}\left[(\hat{a}_{x}^\dagger)^{2}+2\hat{a}_{x}^\dagger\hat{a}_{x}+1+\hat{a}_{x}^{2}\right]\left[(\hat{a}_{x}^\dagger)^{2}+2\hat{a}_{x}^\dagger\hat{a}_{x}+1+\hat{a}_{x}^{2}\right]\nonumber\\
&=\frac{1}{4\gamma_{x}^{2}}\bigg((\hat{a}_{x}^\dagger)^{4}+\hat{a}_{x}^{4}+(\hat{a}_{x}^\dagger)^{2}\hat{a}_{x}^{2}+\hat{a}_{x}^{2}(\hat{a}_{x}^\dagger)^{2}+2(\hat{a}_{x}^\dagger)^{2}+2\hat{a}_{x}^{2}+2(\hat{a}_{x}^\dagger)^{3}\hat{a}_{x}\nonumber\\
&+2\hat{a}_{x}^\dagger\hat{a}_{x}^{3}+2\hat{a}_{x}^{2}\hat{a}_{x}^\dagger\hat{a}_{x}+2\hat{a}_{x}^\dagger\hat{a}_{x}(\hat{a}_{x}^\dagger)^{2}+4\hat{a}_{x}^\dagger\hat{a}_{x}\hat{a}_{x}^\dagger\hat{a}_{x}+4\hat{a}_{x}^\dagger\hat{a}_{x}+1\bigg)
\end{align}
Using the differential calculus employed in Theorem II on page 142 of \cite{Loui}, the above equation (\ref{eq:fourpower}) can be simplified in order to transform all the terms into normal ordered form. 
To this end one replaces $\hat a$ by $z+\partial/\partial z^\ast$ and $\hat a^\dagger$ by $z^\ast$ and applies the expression to the unit operator to arrive at the matrix elements of the normal form of an operator. $\mathcal{N}$ shall denote 
the normal ordering operator. Therefore, e.g., from 
\begin{align}
\mathcal{N}\left\lbrace\left(z+\frac{\partial}{\partial z^\ast}\right)^{2}(z^\ast)^{2}\cdot1\right\rbrace
&=\mathcal{N}\left\lbrace z^{2}(z^\ast)^{2}+4zz^\ast+2\right\rbrace\nonumber\\
&=(z^\ast)^{2}z^{2}+4z^\ast z+2,
\end{align}
it follows that 
\begin{align}
\hat{a}_{x}^{2}(\hat{a}_{x}^\dagger)^{2}&=(\hat{a}_{x}^\dagger)^{2}\hat{a}_{x}^{2}+4\hat{a}_{x}^\dagger\hat{a}_{x}+2
\end{align}
and similarly
\begin{align}
\hat{a}_{x}^{2}\hat{a}_{x}^\dagger\hat{a}_{x}&=\hat{a}^\dagger_{x}\hat{a}_{x}^{3}+2\hat{a}_{x}^{2},
\\
\hat{a}^\dagger_{x}\hat{a}_{x}\left(\hat{a}^\dagger_{x}\right)^{2}&=\left(\hat{a}^\dagger_{x}\right)^{3}\hat{a}_{x}+2\left(\hat{a}^\dagger_{x}\right)^{2},
\\
\hat{a}^\dagger_{x}\hat{a}_{x}\hat{a}^\dagger_{x}\hat{a}_{x}&=\left(\hat{a}^\dagger_{x}\right)^{2}\hat{a}_{x}^{2}+\hat{a}^\dagger_{x}\hat{a}_{x}
\end{align}
for all terms that are not normally ordered already. Hence
\begin{align}
\hat{x}^{4}&=\frac{1}{4\gamma_{x}^{2}}\bigg(\left(\hat{a}^\dagger_{x}\right)^{4}+4\left(\hat{a}^\dagger_{x}\right)^{3}\hat{a}_{x}+6\left(\hat{a}^\dagger_{x}\right)^{2}\hat{a}_{x}^{2}+4\hat{a}^\dagger_{x}\hat{a}_{x}^{3}+\hat{a}_{x}^{4}\nonumber\\
&+6\left[\left(\hat{a}^\dagger_{x}\right)^{2}+2\hat{a}^\dagger_{x}\hat{a}_{x}+\hat{a}_{x}^{2}\right]+3\bigg)
\end{align}
holds for the normal ordered form of the quartic term in the potential and the CS matrix elements of the total potential energy for the system are given by
\be
V_{S}(z_{kx}^\ast,z_{lx})=-\frac{a}{4\gamma_{x}}\left[\left(z_{kx}^\ast+z_{lx}\right)^{2}+1\right]+\frac{b}{16\gamma_{x}^{2}}\left[\left(z_{kx}^\ast+z_{lx}\right)^{4}+6\left(z_{kx}^\ast+z_{lx}\right)^{2}+3\right],
\ee
leading to the corresponding Hamiltonian
\begin{align}
\label{eq:CSDW}
H_{S}(z_{kx}^\ast,z_{lx})&=T_{S}(z_{kx}^\ast,z_{lx})+V_{S}(z_{kx}^\ast,z_{lx})
\end{align}
with the kinetic energy from Eq.\ (\ref{eq:CSkin}).
Now the environmental Hamilton operator
\begin{align}
\hat{H}_{E}&=\sum_{n=1}^{f}\left(\frac{\hat{p}_n^2}{2m}+\frac{m\omega_n^2}{2}\hat{y}_n^2\right)\nonumber\\
&=\sum_{n=1}^{f}-\frac{\gamma_{n}}{4m}\left[\left(\hat{a}^\dagger_{n}\right)^{2}-2\hat{a}^\dagger_{n}\hat{a}_{n}-1+\hat{a}_{n}^{2}\right]+\frac{\gamma_{n}}{4m}\left[\left(\hat{a}^\dagger_{n}\right)^{2}+2\hat{a}^\dagger_{n}\hat{a}_{n}+1+\hat{a}_{n}^{2}\right]\nonumber\\
&=\sum_{n=1}^{f}\omega_{n}\left(\hat{a}^\dagger_{n}\hat{a}_{n}+\frac{1}{2}\right)
\end{align}  
is already normal ordered, leading to the matrix elements
\be
\label{eq:CSenv}
H_{E}(z_{ky_{1}}^\ast,...,z_{ky_{n}}^\ast,z_{ly_{1}},...,z_{ly_{n}})=\sum_{n=1}^{f}\omega_{n}\left(z_{ky_{n}}^\ast z_{ly_{n}}+\frac{1}{2}\right).
\ee
The same holds true for the interaction Hamilton operator
\be 
\hat{H}_{SE}=-\hat{x}\sum_{n=1}^{f}g_{n}\hat{y}_n
\ee
leading to
\be 
\label{eq:CSint}
H_{SE}(z_{kx}^\ast,z_{ky_{1}}^\ast,...,z_{ky_{n}}^\ast,z_{lx},z_{ly_{1}},...z_{ly_{n}})=-\frac{\left(z_{kx}^\ast+z_{lx}\right)}{2\sqrt{\gamma_{x}}}\sum_{n=1}^{f}\frac{g_{n}}{\sqrt{\gamma_{n}}}\left(z_{ky_{n}}^\ast+z_{ly_{n}}\right).
\ee
Finally, the counter-term operator
\be
\hat{V}_C=\hat{x}^{2}\sum_{n=1}^{f}\frac{g_{n}^{2}}{2m\omega_{n}^{2}}
\ee
is just quadratic and leads to
\be
\label{eq:CScount}
V_C(z_{kx}^\ast,z_{lx})=\left[\left(z_{kx}^\ast+z_{lx}\right)^{2}+1\right]\sum_{n=1}^{f}\frac{g_{n}^{2}}{4m\omega_{n}^{2}\gamma_{x}}.
\ee
Therefore, the total normal ordered Hamiltonian is given by the sum of all the terms in Eqs.\ (\ref{eq:CSDW},\ref{eq:CSenv},\ref{eq:CSint},\ref{eq:CScount}) as
\begin{align}
H_{\rm ord}({\bm z}_{k}^\ast,{\bm z}_{l})&=-\frac{\gamma_{x}}{4m_{x}}\left[\left(z_{kx}^\ast-z_{lx}\right)^{2}-1\right]-\frac{a}{4\gamma_{x}}\left[\left(z_{kx}^\ast+z_{lx}\right)^{2}+1\right]\nonumber\\
&+\frac{b}{16\gamma_{x}^{2}}\left[\left(z_{kx}^\ast+z_{lx}\right)^{4}+6\left(z_{kx}^\ast+z_{lx}\right)^{2}+3\right]\nonumber\\
&+\sum_{n=1}^{f}\omega_{n}\left(z_{ky_{n}}^\ast z_{ly_{n}}+\frac{1}{2}\right)\nonumber\\
&-\frac{\left(z_{kx}^\ast+z_{lx}\right)}{2\sqrt{\gamma_{x}}}\sum_{n=1}^{f}\frac{g_{n}}{\sqrt{\gamma_{n}}}\left(z_{ky_{n}}^\ast+z_{ly_{n}}\right)\nonumber\\
&+\left[\left(z_{kx}^\ast+z_{lx}\right)^{2}+1\right]\sum_{n=1}^{f}\frac{g_{n}^{2}}{4m\omega_{n}^{2}\gamma_{x}}.
\end{align}

The complexified classical equation of motion for the displacements $z_{x}$ of the quartic degree of freedom is given by
\begin{align}
{\rm i}\dot{z}_x=\frac{\partial H_{\rm ord}}{\partial z_{x}^\ast}&=-\frac{\gamma_{x}}{2m_{x}}\left(z_{x}^\ast-z_{x}\right)+\frac{b}{4\gamma_{x}^{2}}\left(z_{x}^\ast+z_{x}\right)^{3}-\sum_{n=1}^{f}\frac{g_{n}}{2\sqrt{\gamma_{x}\gamma_{n}}}\left(z_{y_{n}}^\ast+z_{y_{n}}\right)
\nonumber\\
&+\frac{\left(z_{x}^\ast+z_{x}\right)}{2\gamma_{x}}\left[\sum_{n=1}^{f}\frac{g_{n}^{2}}{m\omega_{n}^{2}}-a+\frac{3b}{2\gamma_{x}}\right].
\end{align}
The equations of motion for the displacements $z_{y_{n}}$ of the harmonic (environmental) degrees of freedom are 
\begin{align}
{\rm i}\dot{z}_{y_{n}}=\frac{\partial H_{\rm ord}}{\partial z_{y_{n}}^\ast}=\omega_{n}z_{y_{n}}-\frac{g_{n}}{2\sqrt{\gamma_{x}\gamma_{y_{n}}}}\left(z_{x}^\ast+z_{x}\right).
\end{align}
The equations of motion for the coefficients in the CS expansion are given in (\ref{eq:ccs}).
\end{appendix}


%

\end{document}